\documentclass[12pt]{article}
\usepackage{epsfig}
\usepackage{amssymb}
\usepackage{amsmath}
\usepackage{amsfonts}

\newcommand{\R}{\mathbb{R}}

\newcommand{\Z}{\mathbb{Z}}

\newcommand{\be}{\begin{equation}}
\newcommand{\ee}{\end{equation}}
\newcommand{\bea}{\begin{eqnarray}}
\newcommand{\eea}{\end{eqnarray}}

\newcommand{\kt}{\rangle}
\newcommand{\br}{\langle}

\newcommand{\ed}{\end{document}}

\begin{document}

\title{Pseudo-Hermiticity, ${\cal PT}$-symmetry, \\
and the Metric Operator\footnote{Contributed to the Proceedings of
the 3rd International Workshop on Pseudo-Hermitian Hamiltonians in
Quantum Physics, June 20-22, 2005, Ko\c{c} University, Istanbul,
Turkey.}}
\author{\\
Ali Mostafazadeh
\\
\\
Department of Mathematics, Ko\c{c} University,\\
34450 Sariyer, Istanbul, Turkey\\ amostafazadeh@ku.edu.tr}
\date{ }
\maketitle

\begin{abstract}
The main achievements of Pseudo-Hermitian Quantum Mechanics and
its distinction with the indefinite-metric quantum theories are
reviewed. The issue of the non-uniqueness of the metric operator
and its consequences for defining the observables are discussed. A
systematic perturbative expression for the most general metric
operator is offered and its application for a toy model is
outlined.
\end{abstract}

\section{Introduction}

Most physicists are surprised by being told that certain
non-Hermitian but ${\cal PT}$-symmetric Hamiltonians, such as
$H=p^2+ix^3$, have a purely real spectrum. This was also the case
for the present author who learned about these Hamiltonian in a
seminar given by Miloslav Znojil at Ko\c{c} University in May
2001. Perhaps the most natural question about these Hamiltonian is
whether ${\cal PT}$-symmetry, i.e., the condition $[H,{\cal
PT}]=0$, ensures the reality of their spectrum. It does not take
much effort to see that the answer to this question is a definite
No. Indeed ${\cal PT}$-symmetry is neither necessary nor
sufficient for the reality of the spectrum. Yet it plays a certain
interesting role whose true appreciation has required the use of
the correct mathematical tools.

The search for finding a condition which is both necessary and
sufficient for the reality of the spectrum has led to a notion of
a pseudo-Hermitian operator \cite{p1} that was slightly different
from the one used in the earlier studies particularly in the
context of the indefinite-metric quantum theories and the parallel
mathematical developments \cite{indefinite}.

The approach of \cite{p1} is mainly motivated by the earlier
results on the problem of the geometric phase for non-Hermitian
Hamiltonians \cite{non-Hermitian-ge} and in quantum cosmology
\cite{jmp-98}. According to the definition proposed in \cite{p1},
{\em $H$ is called pseudo-Hermitian, if there exists a Hermitian
and invertible operator $\eta$ satisfying}
    \be
    H^\dagger=\eta H\eta^{-1}.
    \label{ph}
    \ee
In the same article there also appears the notion of an {\em
$\eta$-pseudo-Hermitian} operator, for the case that one fixes a
particular metric operator $\eta$. It is this notion of
$\eta$-pseudo-Hermitian operator that coincides with the older
definition of a pseudo-Hermitian or $J$-Hermitian operator. The
distinction may seem to be quite minute, but in the context of
${\cal PT}$-symmeric QM it has played a most significant
role.\footnote{Indeed the author's lack of knowledge about the
earlier related publications on the indefinite-metric quantum
theories was quite fortunate, for it provided the means to escape
being conditioned by the earlier treatments and allowed for
proposing a notion of pseudo-Hermiticity that proved more useful
in the study of ${\cal PT}$-symmeric systems.}

As is indicated in the title of \cite{p1}, unlike Hermiticity
which is a sufficient condition for the reality of the spectrum,
pseudo-Hermiticity (under some rather general technical
conditions) is a necessary condition. A condition that was both
necessary and sufficient is given in \cite{p2}. It amounts to
supplementing the pseudo-Hermiticity condition with the existence
of a metric operator of the form $\eta_+=O^\dagger O$, \cite{p2}.
Such a metric operator is clearly positive-definite and defines a
positive-definite inner product
$\br\cdot,\cdot\kt:=\br\cdot|\eta_+\cdot\kt$, where
$\br\cdot|\cdot\kt$ is the defining inner product of the Hilbert
space in which $H$ and $\eta_+$ act. This is made more explicit in
\cite{p3} where a clear picture of the role of the antilinear
symmetries such as ${\cal PT}$ is also provided.

\cite{p2} and \cite{p3} also offer other equivalent necessary and
sufficient conditions for the reality of the spectrum of $H$. One
of these is the condition that $H$ may be mapped to a Hermitian
Hamiltonian by a similarity transformation. It was after
publication of \cite{p1,p2,p3} that the author noticed that the
non-Hermitian Hamiltonians having this property are called {\em
quasi-Hermitian}, \cite{quasi}. It is one of the important results
of \cite{p1,p2,p3} that ${\cal PT}$-symmetric Hamiltonians such as
$H=p^2+ix^3$ are quasi-Hermitian. This result cannot be inferred
from those of \cite{quasi}.

An important observation made in \cite{quasi,npb-2002,jmp-2003} is
the non-uniqueness of the metric operator. This problem is
especially important when one deals with the observables of the
theory. The approach of \cite{quasi} to this problem is different
from the one taken in Pseudo-Hermitian QM. In \cite{quasi} the
metric operator is determined by fixing sufficiently many
operators with real spectrum and demanding that they be Hermitian
with respect to the inner product defined by the metric operator.
In contrast in Pseudo-Hermitian QM, one uses the input data which
is the Hamiltonian $H$, to determine the set ${\cal U}^+_H$ of all
possible positive-definite metric operators $\eta_+$. Any two
elements $\eta_{+1}$, $\eta_{+2}$ of ${\cal U}^+_H$ are related
via $\eta_{+2}=A^\dagger\eta_{+1}A$, where $A$ is a symmetry
generator $[A,H]=0$, \cite{npb-2002,jmp-2003}. Each element of
${\cal U}^+_H$ defines a positive-definite inner product and a
complete set of observables that are Hermitian with respect to
this inner product. The arbitrariness in the choice of a complete
set of observables in \cite{quasi} is traded for the arbitrariness
in the choice of an element of ${\cal U}^+_H$. An advantage of the
latter approach is that one can explicitly construct the most
general positive-definite metric operator and the corresponding
observables.

There are two different (currently known) methods of constructing
the most general positive-definite metric operator. the first
method employs the approach pursued in the proof of the spectral
theorems of \cite{p1,p2,p3} and involves constructing a complete
biorthonormal system $\{|\psi_n\kt,|\phi_n\kt\}$ where
$|\psi_n\kt$ and $|\phi_n\kt$ are eigenvectors of $H$ and
$H^\dagger$, respectively. For an explicit application of this
method see \cite{jpa-2004c,p66}. The second method uses the fact
that any positive-definite operator $\eta_+$ has a Hermitian
logarithm, i.e., there is a Hermitian operator $Q$ such that
    \be
    \eta_+=e^{-Q}.
    \label{exp}
    \ee
Inserting this relation in (\ref{ph}), using the
Baker-Campbell-Hausdorff formula,
    \be
    e^{-Q}H\,e^{Q}=H+\sum_{k=1}^\infty \frac{1}{k!}[H,Q]_{_k},
    \label{bch}
    \ee
where $[H,Q]_{_1}:=[H,Q]$ and $[H,Q]_{_{k+1}}:=[[H,Q]_{_k},Q]$ for
all $k\geq 1$, and assuming
    \be
    H=H_0+\epsilon H_1,~~~~~~~~
    Q=\sum_{j=1}^\infty Q_j\epsilon^j,
    \label{expand}
    \ee
where $\epsilon\in\R$, $H_0$ and $Q_j$ are $\epsilon$-independent
Hermitian operators and $H_1$ is an $\epsilon$-independent
anti-Hermitian operator, one obtains and iteratively solves a
system of operator equations for $Q_j$ using perturbation theory.
An explicit application of this method is given in \cite{p64}.

The ${\cal CPT}$-inner product introduced in \cite{bbj-2002} is
just an example of the inner products $\br\cdot,\cdot\kt_+=
\br\cdot|\eta_+\cdot\kt$ where $\eta_+$ has been chosen in a
particular form, namely that ${\cal C}:=\eta_+^{-1}{\cal P}$ is an
involution (squares to one),
\cite{jmp-2003,jpa_2005}.\footnote{Note that, as shown in
\cite{p1}, ${\cal PT}$-symmetric systems such as those considered
in \cite{bbj-2002} are ${\cal P}$-pseudo-Hermitian. Because their
spectrum is real they are also $\eta_+$-pseudo-Hermitian for some
positive-definite metric operator $\eta_+$, \cite{p2,p3}. In view
of Proposition~6 of \cite{p1}, we also have $[\eta_+^{-1}{\cal
P},H]=0$. Note that the condition $(\eta_+^{-1}{\cal P})^2=1$
restricts the choice of $\eta_+$. In general it is not yet clear
if this restriction can always be enforced.} The subsequent
construction of the ${\cal C}$-operator in \cite{bender-jpa-2003}
uses the same approach as in the first method mentioned above. A
variation of the second perturbative method is introduced in
\cite{bender-prd-04}. But because the authors confine their
attention to the ${\cal CPT}$-inner product they miss a large
class of alternative and equivalently admissible positive-definite
metric operators (inner products).

As shown in \cite{p68}, using (\ref{ph}) -- (\ref{expand}) one can
derive
    \be
    \epsilon H_1=\sum_{m=1}^\ell \sum_{k=m}^\ell
    \sum_{j=1}^m\frac{(-1)^j j^k}{k! 2^m}
    \mbox{\small$\left(\!\!\begin{array}{c} m\\j\end{array}\!\!
    \right)$}
    [H_0,Q]_{_k}+{\cal O}(\epsilon^{\ell+1}),
    \label{pert}
    \ee
where $\ell\in\Z^+$, $\left(\!\!\begin{array}{c}
m\\j\end{array}\!\!\right)=\frac{m!}{j!(m-j)!}$, and ${\cal
O}(\epsilon^{\ell})$ stands for terms of order $\ell$ or higher in
powers of $\epsilon$. This equation together with perturbative
expansion of $Q$ given in (\ref{expand}) allows for a systematic
derivation of the operator equations for $Q_j$. For example,
setting $\ell=1,2,3$ in (\ref{pert}) and using (\ref{expand}), we
find
    \be
    \left[H_0,Q_1\right]=-2H_1,~~~~~
    \left[H_0,Q_2\right]= 0,~~~~~
    \left[H_0,Q_3\right]=\frac{1}{12}[H_0,Q_1]_{_3}.
    \label{q1-3}
    \ee
Comparing these relations with those obtained in
\cite{bender-prd-04,p64}, we see that the choice of the ${\cal
CPT}$-inner product corresponds to setting $Q_2=0$ (and putting
certain restrictions on $Q_1$, $Q_3$, etc.) Now, suppose that
$H_0=p^2$ and $H_1=i v(x)$ for a real-valued $v$. Then in the
position representation, Eqs.~(\ref{q1-3}) and their analogs for
$Q_4,Q_5,\cdots$ take the form of a iteratively decoupled system
of non-homogeneous $(1+1)$-dimensional wave equations:
    \be
    (-\partial_x^2+\partial_y^2)Q_j(x,y)=R_j(x,y),
    \label{wave}
    \ee
where $Q_j(x,y)=\br x|Q_j|y\kt$ and $R_j$ only involve
$Q_1,Q_2,\cdots,Q_{j-1}$, \cite{p68}. In view of the fact that the
$(1+1)$-dimensional wave equation has an explicit solution, this
system can be iteratively solved. Note that $R_2=0$, hence
$Q_2(x,y)=f_2(x-y)+g_2(x+y)$ where $f_2$ and $g_2$ are any
complex-valued functions that due to Hermiticity of $Q_2$ satisfy:
$\Re[f_2](-x)=\Re[f_2](x)$, $\Im[g_2](x)=c_2\in\R$, and
$\Im[f_2](-x)=-\Im[f_2](x)-2c_2$.

As an example consider the potential: $v(x)=0$ for $x\notin[-1,1]$
and $v(x)=-{\rm sign}(x)$ for $x\in[-1,1]$. Then the general
solution of (\ref{wave}) for $j=1$ is
    \[Q_1(x,y)=\frac{i}{8}(|x+y+2|+|x+y-2|-2|x+y|-4){\rm sign}(x-y)+
    f_1(x-y)+g_1(x+y),\]
where $f_1$ and $g_1$ have the same properties as $f_2$ and $g_2$.
The appearance of the functions $f_1,g_1,f_2,g_2$ in the
expression for $Q_1$ and $Q_2$ is a manifestation of the
non-uniqueness of the metric operator. Similar constructions
exists for the cases where $H_0=p^2+v_0$ for a real-valued $v_0$.
Further details are given in \cite{p68}.



\end{document}